\documentclass{aa}  

\usepackage{soul}
\usepackage{xcolor}
\usepackage{ulem}
\usepackage{graphicx}
\usepackage{txfonts}
\begin{document} 
   \title{The composition of massive white dwarfs and their dependence on the C-burning modeling.}
    \author{Francisco C. De Ger\'onimo \inst{1,2}
          \and
Marcelo M. Miller Bertolami \inst{3}
          \and
Francisco Plaza\inst{4}
 \and
  Márcio Catelan \inst{1,2}
          }
\institute{Instituto de Astrof\'{\i}sica, Pontificia Universidad Cat\'olica de Chile, Av. Vicuña Mackenna 4860, 7820436 Macul, Santiago,  Chile
\\
\email{francisco.degeronimo@uc.cl}
\and 
Millennium Institute of Astrophysics, Nuncio Monse\~{n}or Sotero Sanz 100, Of. 104, Providencia, Santiago, Chile 
            \\
\email{mcatelan@astro.puc.cl}
\and
Instituto de Astrof\'{\i}sica de La Plata, CONICET-UNLP, La Plata, Argentina
            \\
\email{mmiller@fcaglp.unlp.edu.ar }
\and
Facultad de Ciencias Astron\'omicas y Geof\'{\i}sicas, UNLP, La Plata, Argentina
            \\ 
\email{fran22@fcaglp.unlp.edu.ar}
}

   \date{}

% \abstract{}{}{}{}{} 
% 5 {} token are mandatory
 
  \abstract
   {Recent computations of the interior composition of ultra-massive white dwarfs (WD) have suggested that some white dwarfs could be composed of neon (Ne)-dominated cores. This result is at variance with our previous understanding of the chemical structure of massive white dwarfs, where oxygen is the predominant element. In addition, it is not clear whether some hybrid carbon (C) oxygen (O)-Ne white dwarfs might form when convective boundary mixing is accounted for during the propagation of the C-flame, in the C-burning stage. Both Ne-dominated and hybrid CO-Ne  core would have measurable consequences for asteroseismological studies based on evolutionary models.}
   {In this work we explore in detail to which extent differences in the adopted micro- and macro-physics can explain the different final white dwarf compositions that have been found by different authors. Additionally, we explored the impact of such differences in the cooling times, crystallization and pulsational properties of pulsating WDs.}
   {We perform numerical simulations of the evolution of intermediate massive stars from the zero age main sequence to the white dwarf stage varying the adopted physics in the modeling. In particular, we explore the impact of the intensity of convective boundary mixing during the C-flash, extreme mass-loss rates, and the size of the adopted nuclear networks on the final composition,  age, crystallization and pulsational properties of white dwarfs. }
   {In agreement with previous authors, we find that the inclusion of convective boundary mixing quenches the carbon flame leading to the formation of hybrid CO-Ne cores. Based on the insight coming from 3D hydro-dynamical simulations, we expect that the very slow propagation of the carbon flame will be altered by turbulent entrainment affecting the inward propagation of the flame.   Also, we find that Ne-dominated chemical profiles of massive WDs recently reported appear in their modeling  due to the overlooking of a key nuclear reaction. We find that the inaccuracies in the chemical composition of ultra-massive white dwarfs recently reported lead to differences of 10\% in the cooling times and degree of crystallization and about 8\% in the period spacing of the models once they reach the ZZ Ceti instability strip. }
   {}

     \maketitle
%-------------------------------------------------------------------

\section{Introduction}

White dwarf stars (WDs) are the most common end product of stellar evolution. Stars with initial masses below $M_{\rm ini}\sim 8-10 \,M_{\odot}$ are known to end their lives as WDs \citep{2008ARA&A..46..157W, 2010A&ARv..18..471A}. 
These stars are valuable tools for a wide variety of applications in astrophysics, from being age and distance indicators \citep{2016NewAR..72....1G, 2017ApJ...837..162K} to laboratories to test extreme physical conditions \citep{2009ApJ...693L...6W,2019Natur.565..202T,2021A&A...649L...7C}.
While most of these stars have gone through He-core burning and the asymptotic giant branch (AGB) leading to the formation of carbon (C)-oxygen (O) cores, some other evolutionary paths are possible. 
Low-mass WDs ($M\lesssim 0.45 \,M_\odot$),
 are formed by stars in binary systems or with primordially high initial He abundances, that avoid He-core burning and lose their H-rich envelopes already on the first giant branch (RGB), thus harboring He-cores \citep{2010A&ARv..18..471A}. Stars at the boundary between intermediate- and high-mass stars ($M_{\rm ini}\sim 8-10 \,M_{\odot}$) are particularly interesting in this respect. These stars reach temperatures that are high enough to ignite their CO cores under degenerate conditions \citep{1994ApJ...434..306G},  and then evolve into the so-called super AGB (SAGB) phase.
 Although the basic global properties of SAGB stars are relatively well determined, the final chemical structure  left at the end of the phase is still a matter of debate.
 Classic works by \cite{1994ApJ...434..306G}, \cite{ 1996ApJ...460..489R}, \cite{1997ApJ...485..765G} and \cite{1997ApJ...489..772I} showed that the C-flash and subsequent C-burning leads to a oxygen-neon (ONe) core and, consequently to an ONe WD \citep[see][and references therein]{2006A&A...448..717S, 2007A&A...476..893S, 2010A&A...512A..10S, 2019A&A...625A..87C} or an electron-capture supernova \citep{2013ApJ...771L..12T}, depending on the intensity of winds.  \cite{2013ApJ...772...37D} claimed that the existence of even a small amount of convective boundary mixing (CBM) below the convective zone generated by the C-flash leads to the quenching of the C-flame and the formation of a hybrid CO-Ne degenerate core. Recently, from the computation of fully evolutionary sequences, \cite{2018MNRAS.480.1547L} found models of WDs with masses higher than $1.15\,M_{\odot}$ harboring Ne-dominated cores (Ne-O-Mg WDs), and hybrid CO-Ne WD models with a composition significantly different from that obtained by previous works.
 
As is well known, the properties of the non-radial $g$-mode pulsations of pulsating WDs are tightly connected to the chemical structure \citep[see, for example][]{2021A&A...646A..30A, 2017A&A...599A..21D}.
 Thus, the precise internal chemical profile of SAGB stars is of the utmost importance for the proper employment of the asteroseismological techniques based on fully evolutionary models applied to ultra-massive pulsating WDs  \citep[see][and references therein]{2019A&ARv..27....7C, 2019A&A...621A.100D}.  In this regard, the discrepancy seen in the abundances reported by different authors may alter the results derived not only from asteroseismological analysis, but also from cooling studies. 
   It is not unthinkable that the disparity in the compositions derived by previous authors could be due to the differences in the adopted physical ingredients. In particular, extreme assumptions about the micro- and macro-physics, particularly in regard to overshooting, stellar winds, and/or key nuclear reaction reaction rates, could plausibly lead to the (spurious) predicted formation of exotic objects.
  
 In this paper we assess how the different physical ingredients adopted during C-burning affect the final composition of massive WDs. We specifically focus on the impact of mass-loss rates, size of the nuclear networks, and the intensity  of convective boundary mixing. The paper is organized as follows, in section ~\ref{sect:chemicalSAGB} we provide a brief description of the typical chemical structures on the SAGB. In section~\ref{sect:impact} we explore the impact of CBM, extreme winds, and the size of the nuclear network, and discuss their consequences for the final composition of the WD model.   Section \ref{sect:consequence} is devoted to explore the impact of the different chemical compositions on the cooling times, crystallization and pulsational properties of DA WDs. Finally, in section~\ref{sect:concl}, we close the article with some concluding remarks.

        \begin{figure}
  \includegraphics[width=0.75\columnwidth,angle=270]{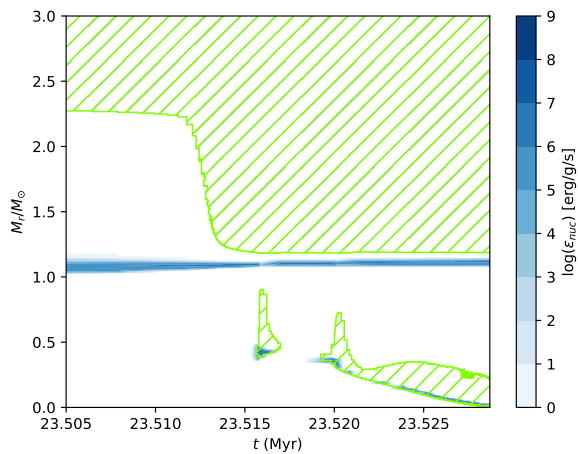}
   \caption{Kippenhahn diagram during the C-burning phase for a  model of $M_{\rm ZAMS}= 10 \, M_{\odot}$ with $Z=0.02$, in absence of extra-mixing processes.
   The green areas are the convective zones, while the blue scale represents the net energy from nuclear reactions. In this case, the flame successfully reaches the center, leaving a pure ONe core.}
  \label{fig:kipp-0}
\end{figure}

     \begin{figure}
  \includegraphics[width=0.75\columnwidth,angle=270]{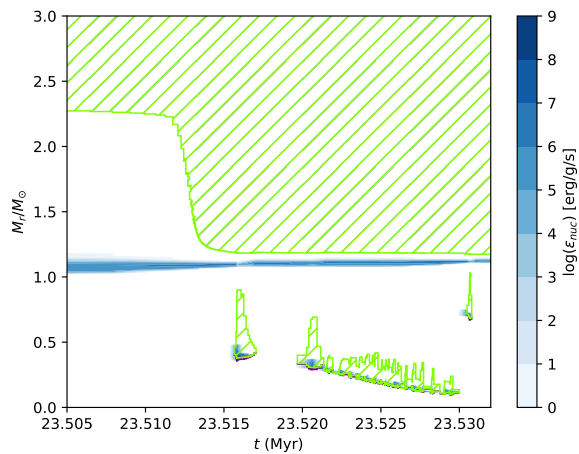}
   \caption{Kippenhahn diagram for the same $10\,M_{\odot}$ model as in Fig. ~\ref{fig:kipp-0}, but when diffusive
   overshooting is allowed at the bottom of the C-burning shell with overshooting parameter $f=0.007$ (purple area). In this case, 
   the flame stops before reaching the center thus leaving a unaltered CO core surrounded by the ashes of C-burning.}
  \label{fig:kipp-07}
\end{figure}

%--------------------------------------------------------------------

 \section{Chemical signatures of SAGB stars}
 \label{sect:chemicalSAGB}
The evolutionary behavior of SAGB progenitors before carbon
ignition is very similar to that of intermediate-mass stars that end
up as CO WDs, and is well documented in a myriad of papers \citep{1994ApJ...434..306G, 2006A&A...448..717S, 2010A&A...512A..10S}. For
the sake of completeness, and to set the stage for the study of the
C-burning phase in the next sections, we provide here only a brief
description of the previous evolutionary stages. 

During the main sequence
SAGB progenitors undergo hydrogen burning via the CNO cycle inside
a convective core. Immediately after central hydrogen (H) exhaustion, the H-burning sets in a
shell  and the helium (He)-core contracts,  with the consequent increase of the
temperature. During this brief red giant phase, the star experiences the
first dredge up, polluting the envelope with material previously
processed by H-burning. Once temperatures exceed $\sim 2\times 10^8$ K, helium
ignites in the core under non-degenerate conditions, and burns  in
a convective core. It is known that CBM
during both core H- and He-burning leads to larger convective cores \citep[e.g.][]{1985ApJ...294L..31C, 2012A&A...537A.146E,2016MNRAS.456.3866C, 2020MNRAS.493.4748W}, than those predicted by a strict
application of the Schwarzschild criterion \citep{1906WisGo.195...41S}.  In our work, CBM during both core H- and He-burning was not included. This choice was motivated by our desire to more directly compare our simulations with those presented by \citet{2018MNRAS.480.1547L}, who did not take into account CBM during these earlier phases. The main impact
of CBM in these stages is to decrease the initial mass required for a
progenitor star to reach C-ignition by about $2\, M_\odot$ (the minimal mass 
for C-ignition shifting from $M_{\rm ZAMS}\sim 8.5 \,M_\odot$ to  
$\sim 6.5 \,M_\odot$ for standard assumptions in CBM, where ZAMS stands for the zero-age main sequence). After the
end of core He burning, these objects will experience the second
dredge up, as do their lower mass counterparts. The second dredge up reduces
the mass of the contracting CO core, preventing the ignition of carbon
in non-degenerate conditions. Finally, the
degenerate CO core eventually reaches a critical mass of $M_{\rm CO}\sim 1.05\,
M_\odot$, and carbon ignites. % off-centered in degenerate conditions
%leading to a carbon flash.
%The SAGB phase is the last evolutionary stage in which stars with initial masses above $\sim 8-10M_{\odot}$, ends the building of its final chemical structure.
% are the result of the complex interplay between the nuclear reactions and their energy released together with the efficiency of the mixing processes. 
%In Fig.~\ref{fig:perfil-bien} we show  the expected chemical profile in terms of the mass fraction, for a  model with initial mass 10$M_{\odot}$ computed from ZAMS to the 40th thermal pulse in the SAGB.

The detailed chemical structure of SAGB stars is expected to leave its imprint in the interiors of ultra-massive WDs. These features, are the signatures of the different physical processes that are operative during the whole evolution in such a way that different regions of the chemical profile can be the trace of individual processes. 
In particular, the expected core chemical structure found at the SAGB phase for massive progenitors is the result of the evolution through the C-burning stage.
Under the assumption of strict Schwarzschild criterion \citep{1906WisGo.195...41S} and no additional mixing processes the development of carbon burning is characterized by two clearly different stages \citep{1994ApJ...434..306G, 2006A&A...448..717S}. The first corresponds to the ignition of C at the point of maximum temperature inside the partially degenerate CO core, inducing a thermal runaway  which is referred to as the carbon flash. The sudden energy injection by the C-flash leads to the development of a convective zone which extends outwards from the point of maximum temperature. The second stage corresponds to the development of a flame which propagates all the way to the center and transforms the CO core into an ONe core \citep{1994ApJ...434..306G,  2006A&A...448..717S}.
%-------------------------------------------------------

%------------------------------------------------------------
In Fig.~\ref{fig:kipp-0} we show the Kippenhahn diagram for a 10$\,M_{\odot}$ and $Z=0.02$ model for which no CBM was adopted. 
 During the C-flash, a convective region develops that lasts a few thousand years. The next convective region develops during the C-burning and remains until the C-flame turns off at the center of the star.
 However, \cite{2013ApJ...772...37D} showed that the inclusion of even a small amount of CBM during the propagation of the C-flame removes the necessary conditions for  further propagation of the C-flame, inducing an anticipated halt of the carbon flame.  In this context,  if a reasonable amount of CBM is considered, the  C-burning phase proceeds as a main flash and subsequent sub-flashes of less intensity and neither the flame nor the mixing reaches the center \citep[for a detailed discussion see][ and references therein]{2015ApJ...807..184F}.  This is seen in Fig.~\ref{fig:kipp-07}, where the Kippenhahn diagram is displayed for the same model but adopting diffusive overshooting with $f=0.007$ during the C-burning phase \citep[see Section 3 from][]{2013ApJ...772...37D}.
Thus, those objects should show a  marked contrast in the  chemical structure at the end of the C-burning phase, which is expected to  be composed of a pure unaltered CO-core, produced during the core helium burning (CHeB) phase, surrounded by an ONe mantle. 

Finally, for ultramassive WD progenitors, the aftermath of the evolution during the thermally unstable phase is, first, the build-up of the most external part of the core, where the He-burning shell was active and, second, a strong reduction of the hydrogen and helium content down due to mass loss.

     \begin{figure}
  \includegraphics[width=1\columnwidth]{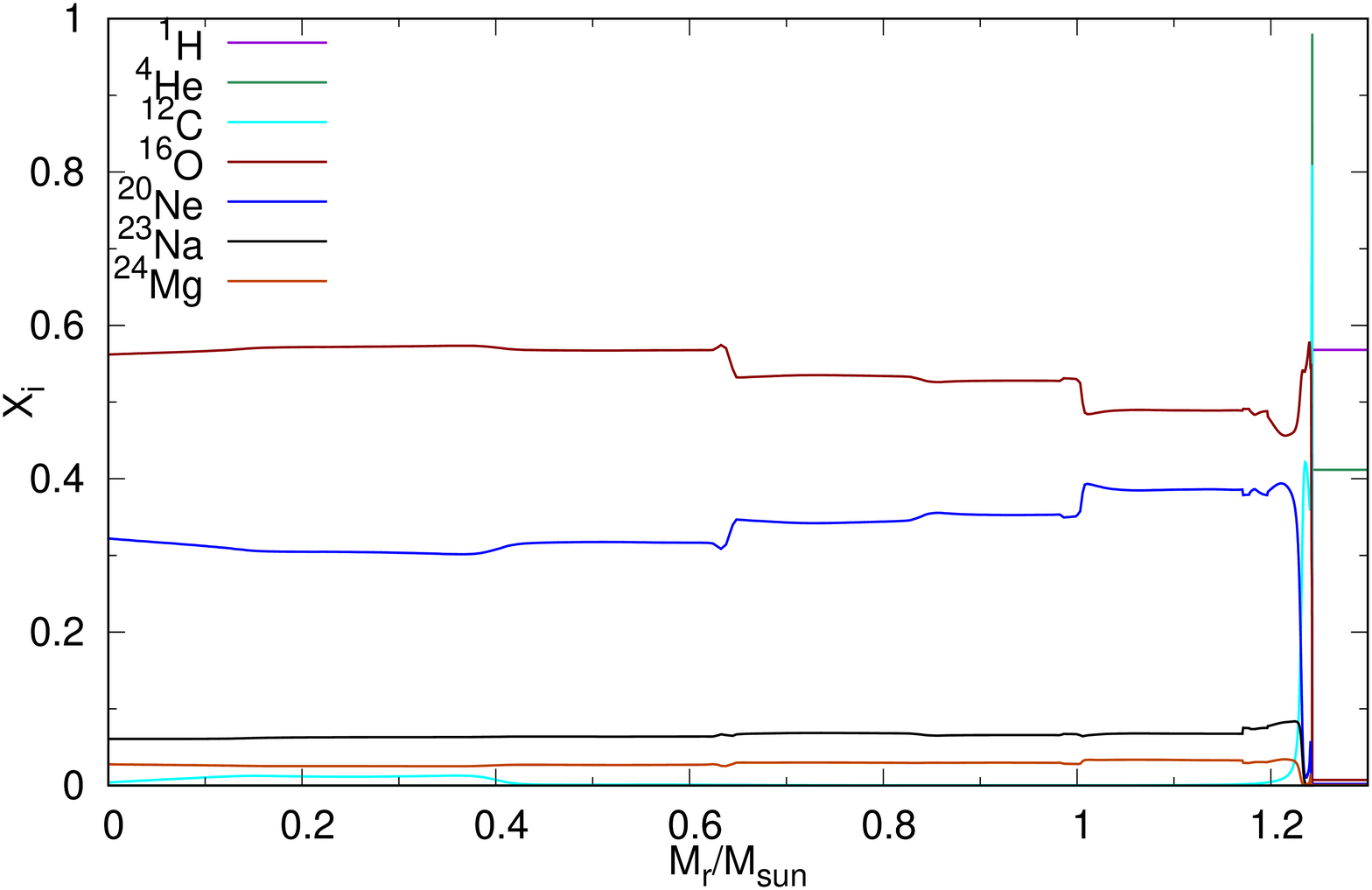}
   \caption{Chemical profiles for the most abundant species at the end of the C-burning phase for a  model computed with $M_{\rm ZAMS}=11\,M_{\odot}$  and $f=0$ during the C-burning phase. The core is composed of a mixture of $^{16}{\rm O}$ and $^{20}{\rm Ne}$ with traces of $^{12}{\rm C}$, $^{23}{\rm Na}$ and $^{24}{\rm Mg}$.}
  \label{fig:f0}
\end{figure}
  
 \begin{figure*}
  \includegraphics[width=\textwidth]{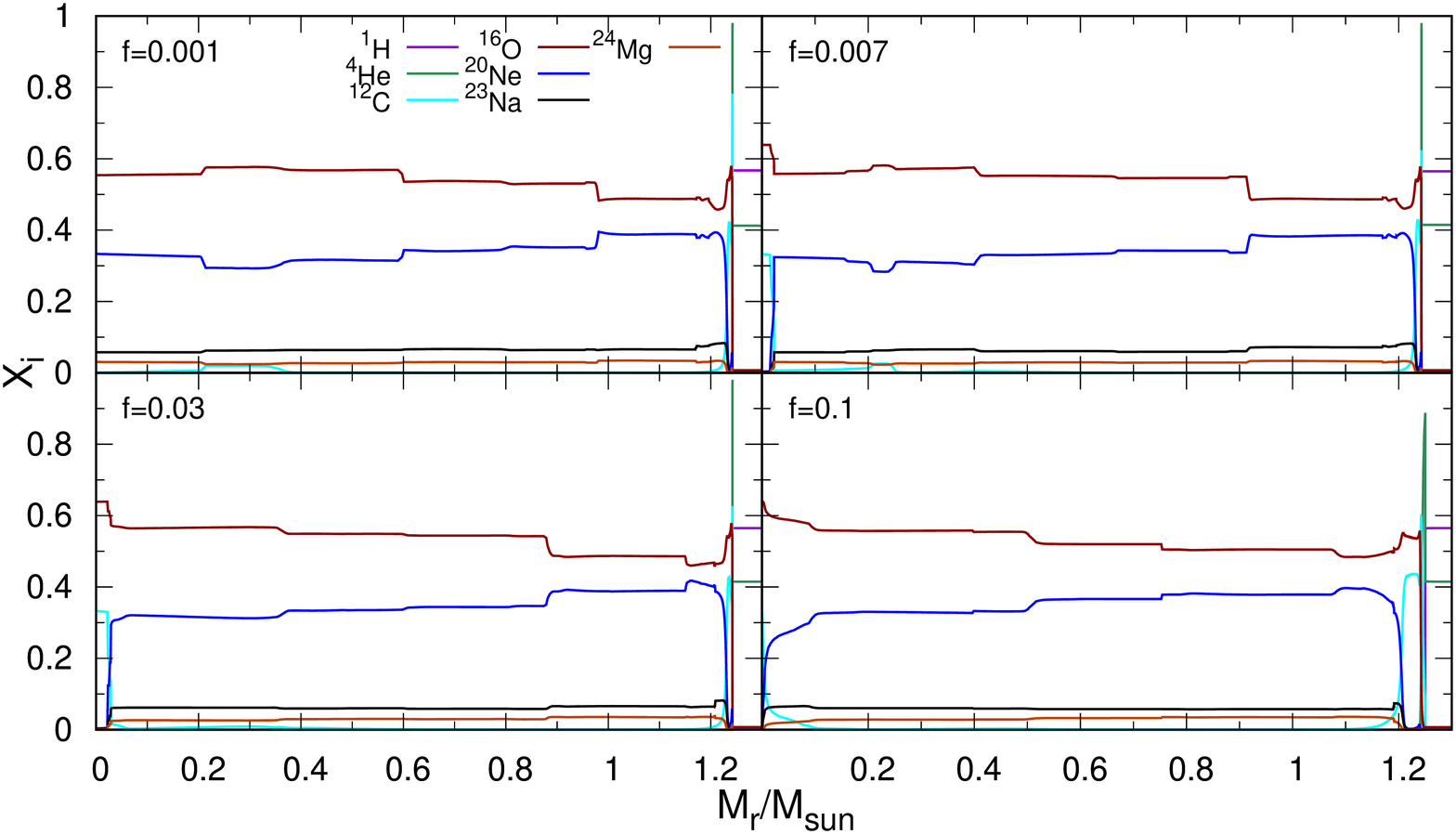}
    \caption{Chemical structures for the $11 \, M_{\odot}$ models at the end of the C-burning phase,  computed assuming overshooting parameters $f=~0.001,~0.007,~ 0.03, ~0.1$. For moderate-to-high overshooting, the C-flame halts before reaching the center of the star. The enhanced mixing induced in the $f=0.1$ model strongly reduces the size of the pure CO core.}
  \label{fig:f-todos}
\end{figure*}

 \section{Impact of physical assumptions on the final composition of the WD}
 \label{sect:impact}
 All the models presented in this work were computed with the stellar evolution code Modules for Experiments in Stellar Astrophysics ({\tt MESA}) versions 8445 and 12778 \citep{Paxton2011, Paxton2013, Paxton2015, Paxton2018, Paxton2019}. Most of the adopted input physics corresponds to the default options that are described in detail in \cite{Paxton2011, Paxton2013, Paxton2015, Paxton2018, Paxton2019}, and will thus not be repeated here. Exceptions include mixing at the border of convective regions, mass loss, and nuclear reaction networks, which will be addressed in the following subsections.

 As mentioned in the Introduction, the existence of Ne-dominated and hybrid CO-Ne massive WDs is relevant both for asteroseismological studies and for our understanding of the cooling timescales of massive WDs. Given that it has been shown that no significant differences in the predictions for SAGB stars arise simply from the use of different stellar evolution codes \citep{2010MNRAS.401.1453D}, it is safe to assume that the predicted existence of Ne-dominated or hybrid CO-Ne WDs  depends on the adopted physics. To analyze how robust this prediction is, we will here analyze what the key physical assumptions are that can lead to these results. 
  While the main difference in the physical ingredients in the computation of hybrid CO-Ne WDs and pure ONe WDs relies in the inclusion of CBM during the propagation of the C-flame, the main differences between the predicted formation of Ne-dominated WDs \citep{2018MNRAS.480.1547L} and the different results previously reached by other authors can be traced back to the extension of the CBM  region, the adopted mass loss rates, and, finally, the size of the nuclear reaction network. In what follows, we discuss each of these individually.

  \subsection{Impact of convective boundary mixing}
  Convective boundary mixing has the potential to affect the final composition of the core by affecting the relative proportions of C and O during the C-flash,  and also by allowing material to be mixed down into deeper regions of the star, effectively affecting the mean temperatures at which carbon is burnt. Unfortunately, CBM is one of the least understood phenomena in stellar evolution \citep{2019MNRAS.484.4645C}. It is usually parameterized and must be calibrated with observations of different stellar objects \citep[e.g.,][]{2016A&A...588A..25M}, and no calibration exists for the convective regions driven by C-burning. In this context, we will estimate the expected range of CBM parameters and, then, investigate how much the final composition can change as a consequence of this.

  \subsubsection{Estimating the efficiency of CBM}
 
    As shown by \citet{2009A&A...497..463S} and \citet{2013ApJ...772...37D}, the inclusion of mixing processes below the flame has the potential to quench the latter flame, leading  to the formation of a degenerate CO core surrounded by a mantle composed mostly of O and Ne. It is not clear, however, to which degree the different assumptions about CBM is solely responsible for the great disparity in the final chemical compositions of previous works \citep{2007A&A...476..893S, 2013ApJ...772...37D, 2018MNRAS.480.1547L}, or  whether other ingredients involved in the computations (such as nuclear reaction rates and networks) also play a significant role. Furthermore, it is not clear from previous works what is the underlying physical reason for the different assumptions of the CBM. Given the aforementioned relevance of CBM assumptions for the final composition of the WD, it would be highly desirable to know which values are acceptable.
    
    Simple estimations of the adiabatic penetration of convective elements  into the stable layers can be done by means of the following expression:
    \begin{equation}
        \frac{\bar{v}_{\rm MLT}^2}{2}=\int_{r_{\rm FCB}}^{r_{\rm CBM}}   g(r) \left [ \frac{\rho(r)-\rho_e(r)}{\rho_e(r)}\right] dr,
    \end{equation}
    where $\bar{v}_{\rm MLT}$ is the mean value of the convective velocities close to the formal convective boundary (FCB), as estimated by the mixing length theory (MLT); $\rho_e$ is the density of the convective elements moving adiabatically from the convective boundary; and $\rho$ and $g$ are the local values of the density and gravity acceleration at distance $r$ from the center of the star. By applying this estimation to an initially $10 \, M_\odot$ model during the peak of the C-flash, computed without CBM, we obtain that the distance penetrated by convective elements at the lower convective boundary is  $d_{\rm pen}\simeq 2.5\times 10^{-4} H_P$ ---with $H_P$ the value of the pressure scale height at the formal convective boundary, and $H_P=1.9\times 10^8$~cm. This is a very small value, with no practical consequences for the development of the C-flash.
    
    However, during a thermal runaway phase, CBM can be caused by turbulent entrainment \citep[see][]{2007ApJ...667..448M,2015A&A...580A..61V}. In the case of turbulent entrainment, turbulence diffuses into the nearby stable region, and leads to a progressive advance of the border of the turbulent region. According to \citet{2007ApJ...667..448M}, the velocity $v_E$  at which the turbulent border gains ground on the stable layer is 
    $v_E=\sigma\, A\, {\rm Ri_{\rm B}}^{-n}$, 
    where ${\rm Ri_{\rm B}}$ is the bulk Richardson number, $\sigma$ is the rms velocity at the turbulent border, and $n$ and $A$ are two free parameters to be calibrated with numerical experiments. The bulk Richardson number is
a measure of the stiffness of the boundary region, and is defined as
${\rm Ri_{\rm B}}=\Delta b\, L/\sigma^2$, where L is the mixing length scale and $\Delta b$ is the relative buoyancy jump. The latter is defined as
\begin{equation}
\Delta b=\int_{r_0}^{r_1} N^2 dr,
\label{eq:b_jump}
\end{equation}
where $r_0$ and $r_1$ are the bottom and upper parts of the transition region and $N$ is the adiabatic buoyancy frequency. We can then estimate the distance traveled by the lower convective boundary during the carbon flash as
\begin{equation}
d_E=|r_{\rm FCB}-r_{\rm CBM}|=v_E\, \tau_{\rm C-flash}.
\end{equation}  
where $ \tau_{\rm C-flash}$ is estimated as the time required for C-burning to drop by one order of magnitude.
 From the numerical experiments of \cite{2007ApJ...667..448M}, it results that $A\simeq 1$ and $n\simeq 1$, so for practical purposes we can estimate $v_E\simeq \bar{v}_{\rm MLT}/ {\rm Ri_{\rm B}}$ then, approximating $\sigma\simeq \bar{v}_{\rm MLT}$, equating $L$ to the mixing length and setting $r_0=r_{\rm FCB}$, $r_1=r_{\rm CBM}$ in eq. \ref{eq:b_jump} the distance entrained by turbulent border during the carbon flash can be estimated as \citep{2020MNRAS.493.4748W}
\begin{equation}
d_E\simeq \frac{\tau_{\rm C-flash} \bar{v}_{\rm MLT}^3}{\Delta b L_{\rm MLT}}.
\end{equation}  
Using the values from the same $10 \, M_\odot$ model during the peak of the C-flash as before, we have $\tau_{\rm C-flash}\simeq 28$~yr, $L_{\rm MLT}\simeq 3.4\times 10^8$~cm, and  $\bar{v}_{\rm MLT}\simeq 103\,600$~cm/s implying
$d_E\simeq 0.15 H_P$. This value is surprisingly close to the usual size of the CBM region in upper main sequence stars, and is not completely negligible. We can conclude that the possible impacts of CBM during the C-flash stage are worth exploring. 

It is worth mentioning, however, that despite its significant value, the estimate presented in the previous paragraph is several times smaller than the CBM extension adopted by \cite{2018MNRAS.480.1547L}. The latter authors adopted an overshooting parameter value of $f=0.1$; by looking at our own $11 \, M_\odot$ sequence with $f=0.1$ (see next sections), this corresponds to a formal overshooting region of size $\delta r_{\rm OV} \simeq 0.003 R_\odot$, or 1.2$H_P$. Even if we restrict ourselves to the region where material is actually being mixed during the C-flash, the size of the overshooting zone in that model is $\delta r_{\rm OV} \simeq 0.002 R_\odot$  (0.8$H_P$).

According to the previous results in the literature,  even more important is the intensity of CBM during the flame propagation \citep{2009A&A...497..463S, 2013ApJ...772...37D}. Interestingly, from the same $10 \, M_\odot$ model as before but during the peak of the C-flame we get $\bar{v}_{\rm MLT}=48\,000$~cm/s. The deepening of the convective zone in this model happens at a rate of about $v_{{\rm flame}}\sim 10^{-3}$~cm/s. Given that for any reasonable estimation of ${\rm Ri_{\rm B}}$ we get ${\rm Ri_{\rm B}<10^7}$, this argument suggests that $v_e\gg v_{\rm flame}$. Thus,  the timescale for the inwards propagation of the flame will be strongly altered if turbulent entrainment is included. We conclude that quenching of the C-flame is highly likely when turbulent entrainment is included, in agreement with the suggestions by \citet{2013ApJ...772...37D}.

\begin{figure}
  \includegraphics[width=1\columnwidth]{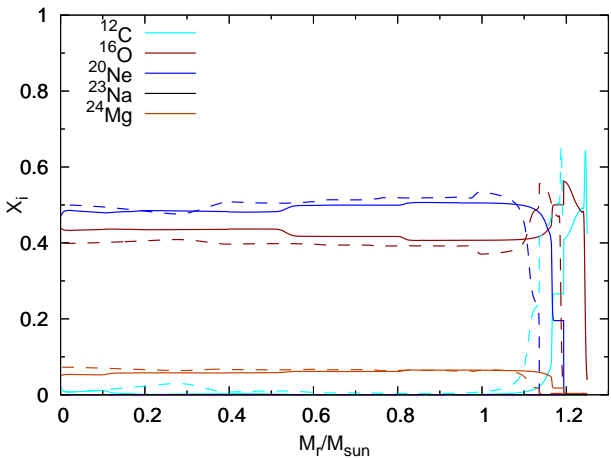}
    \caption{Chemical profiles for the model computed with the nuclear network {\tt co\_burn\_plus.net} and overshooting parameter $f=0.1$ and  $f=0.$ (solid and dashed lines, respectively). }
  \label{fig:ov-lauffer}
\end{figure}

  \subsubsection{Impact of CBM on evolutionary sequences}
\label{sect:CBM}

We computed evolutionary sequences with $M_{\rm ZAMS}=11 \, M_{\odot}$ and  initial composition $Z=0.02$, from the ZAMS through the  central hydrogen and helium burning stages, the RGB and finally to the SAGB stage. A rather large nuclear reaction network (namely {\tt sagb\_NeNa\_MgAl.net}) was adopted. This includes 29 species, from neutrons to $^{27}Al$, and which incorporates several carbon burning reactions and both proton and alpha captures by $^{12}$C nuclei. No mass loss was included, as winds before the thermally pulsing SAGB phase are almost negligible \citep[see, for example, table 2 from][]{2007A&A...476..893S}.  In order to test the impact of CBM on the final composition  of the core, and in line with the discussion of the previous section, different values of the overshooting parameter $f$ at the bottom of the C-burning convective zones were considered, namely $f=0, ~0.001,~ 0.007,~ 0.03, ~0.1$.

In Figs. \ref{fig:f0} and \ref{fig:f-todos} we show the expected internal structure (i.e., of the innermost $1.3 \, M_{\odot}$)  at the moment that C-burning stops, for  models computed with different values of the parameter $f$. It is seen that for $f=0$ and 0.001 the flame successfully reaches the center of the star, leaving a pure ONe core. 
On the other hand, the inclusion of moderate-to-high overshooting results in the   premature halt of the C-flame, with the consequent formation of 
hybrid cores composed of an inner unaltered CO core surrounded by an ONe mantle.
The size of the unaltered CO-core is determined by the location of the quenching of the flame, and also depends on the efficiency of the extra-mixing. For the $f=0.1$ case, Fig.~\ref{fig:f-todos} shows that this assumption strongly reduces the  CO-core size and creates a new chemical interface where $^{12}{\rm C}$, $^{16}$O and $^{20}$Ne coexist. This reduction is a consequence of the large extension of the CBM below the convective zone generated after the main C-flash.  Thus, although the existence of CBM still leads to the quenching of the C-flame, $^{20}$Ne was already carried close to the center by the main C-flash. In none of the cases explored  were we able to produce Ne-dominated cores. We can safely conclude that the inclusion of enhanced overshooting is not responsible for the formation of the NeO-core\footnote{The name NeO(ONe)-core is adopted when the core is composed mostly by $^{20}$Ne($^{16}$O).}  WDs found by \citet{2018MNRAS.480.1547L}.

To test the robustness of this conclusion, we analyzed it under the physical assumptions established by \citet{2018MNRAS.480.1547L}. To do this, we computed the evolution of models with initial mass 11$M_{\odot}$,  adopting the same nuclear network  as in \citet{2018MNRAS.480.1547L} i. e. {\tt co\_burn\_plus.net} (see section \ref{sect:network} for details on the nuclear networks), from the ZAMS to the end of the C-burning phase, considering enhanced mass loss during the RGB and AGB, using MESA r8445. 
 For accounting for CBM processes during C-burning, diffusive overshooting was adopted along the whole evolution for  metal burning regions ("{\tt burn\_z}" option in {\tt MESA}), and thermohaline mixing was allowed.
 In Fig.  \ref{fig:ov-lauffer} we show the expected chemical profiles for models with overshooting parameter $f=0.1$ and $f=0$ (solid lines and dashed lines, respectively). 
Two distinctive features arise: the formation of Ne-dominated cores (see next section) and the absence of the hybrid core  when overshooting is adopted. 
The latter phenomenon results is a consequence of both enhanced extra-mixing, resulting from the adoption of $f=0.1$, and the presence of thermohaline mixing,  which eliminated the small, pure CO core that would otherwise have been expected.
 As pointed out previously, it is seen that the inclusion of enhanced overshooting does not alter significantly the distribution of elements in the core.

 From the same figure, a notable difference in the size of the NeO core is perceived. We find that, when accounting for overshooting in the metal burning regions with the parameter "{\tt burn\_z}", the burning of $^{14}$N in the CHeB phase induces an overshooting region that moves the border of the core farther away by an additional $\sim$0.22 $M_{\odot}$, thus affecting the post-CHeB evolution.
%during  the beginning of the CHeB phase, the development of an overshooting region . %There we  show the chemical structure for the inner part of the model, energy released by nuclear reactions and mixing regions, for two consecutive models during CHeB: at the moment and post the development of a overshooting region (dotted and solid lines respectively). 
%By exploring the nuclear reactions that are operative during this phase, we found a 
% that the energy released during the burning of $^{14}$N induces an overshooting region, when it is included under the condition  "{\tt burn\_z}". This moves the border of the core farther away, by an additional $\sim$0.22 $M_{\odot}$, thus affecting the post-CHeB evolution. 
For a  correct assessment of the  overshooting process during the carbon burning phase, it should be turned on at the beginning of said phase. 

\subsection{Impact of extreme mass loss on the early SAGB}
\label{subsec:massloss}
The thermally unstable phase post C-burning is one of the most time demanding phases when computing the evolution of SAGB stars. During this phase, the star expands and mass loss becomes important. As low mass stars in the AGB, SAGB stars also experiences recurrent thermal instabilities in the shell where He is burning known as thermal pulses. Such instabilities are characterized by a sudden injection of energy in a relatively small time lapse. 
To avoid the numerical complications connected with the computation of the thermal pulses, \citet{2018MNRAS.480.1547L} adopted values of mass loss rates during AGB higher than recommended. Particularly, the authors implemented the  formulae for AGB mass loss from \cite{1995A&A...297..727B}, adopting a large value of  $\eta_{Blocker}=10$, compared with the more usual $\sim 0.1$, which is reflected as an enhanced wind.
This results in extreme mass loss  previous to the C-ignition, leading  not only to the complete removal of the H-rich envelope, but also to the removal of most of the He-rich mantle before the development of the C-flash. 
Therefore, the H-burning shell is completely extinguished and something similar happens with  the He-burning shell, which finally results in a strong alteration of the post-AGB evolution, affecting  both the path on the Hertzsprung-Russell diagram and the chemical structure and thermal stratification of the CO core. As a consequence,  the ignition of C, which takes place at a  position closer to the center than in the absence of mass loss, occurs when the star has lost almost 90\% of its mass.
  Additionally, we noted that the He content added by \cite{2018MNRAS.480.1547L} to  their $M_{\rm WD}\geq 1.132 M_{\odot}$ models to replace the  layers lost  by artificially strong winds is inconsistent with the predictions of detailed SAGB models, being almost 10 times higher for a $M_{\rm WD}=1.22 \, M_{\odot}$ model %$\sim 8\times 10^{-5}M_{\rm WD}$. 
\cite[][]{2010A&A...512A..10S}. This is expected to have a non-negligible impact on the cooling times and crystallization of WDs (see the next sections).
Finally, we found that the high mass loss rate adopted neither is the cause of the Ne-enhanced core compositions found by the authors.

  \label{sect:ev-seq}

   \begin{figure}
  \includegraphics[width=1\columnwidth]{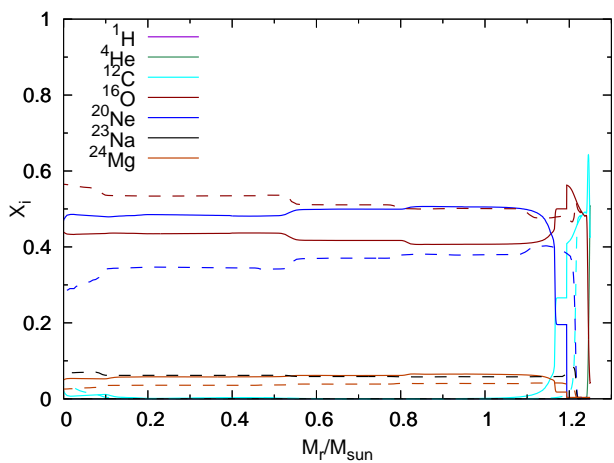}
    \caption{Chemical structures for 11$M_{\odot}$ models at the end of the C-burning phase with an enhanced mass loss and nuclear networks {\tt co\_burn\_plus.net} (solid line) and {\tt sagb\_NeNa\_MgAl.net} (dashed line).}
  \label{fig:rednuc}
\end{figure}
   
\begin{figure}
  \includegraphics[width=1\columnwidth]{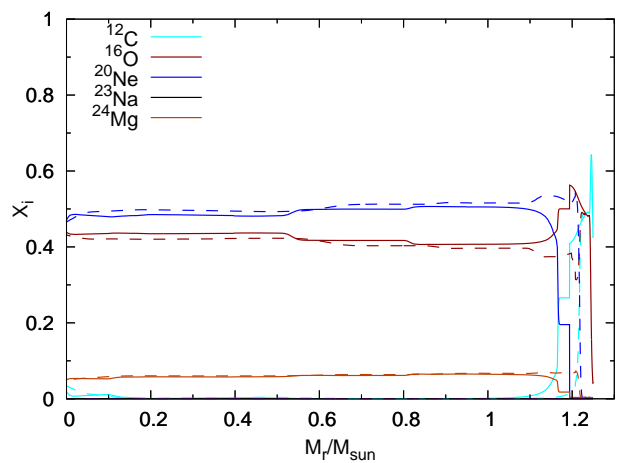}
    \caption{Chemical profiles for 11$M_{\odot}$ models considering the  nuclear network {\tt  co\_burn\_plus.net} (solid line) and the nuclear network {\tt sagb\_NeNa\_MgAl.net} without the $^{12}{\rm C}(^{12}{\rm C},p)^{23}{\rm Na}$ reaction (dashed line).}
  \label{fig:no23na}
\end{figure}

 \subsection{Impact of nuclear reaction rates}
\label{sect:network}
Previous works on the modeling of SAGB stars have adopted nuclear reaction networks of very different sizes, from 52 species in \cite{2006A&A...448..717S} to 16 species in  \citet{ 2018MNRAS.480.1547L}. Given the large differences in the sizes of the nuclear networks, it is expected that some differences in the  chemical profiles may arise.
 In this section, we explore the effects of the assumption of different nuclear reaction networks on the shape of the chemical profiles expected at the end of the SAGB phase.

 We computed models with the same input physics as in \cite{2018MNRAS.480.1547L}, including thermohaline mixing. We also analyzed the impact of two different reaction networks, namely {\tt co\_burn\_plus.net}, which was used by \citet{2018MNRAS.480.1547L} and follows 16 species, and {\tt sagb\_NeNa\_MgAl.net}, which includes 29 species in total. The choice to incorporate {\tt sagb\_NeNa\_MgAl.net} into our analysis is motivated by the fact that it includes the $^{23}$Na isotope (unlike {\tt co\_burn\_plus.net}), with potentially important consequences for the resulting chemical profile. Indeed, the primary reaction $^{12}{\rm C}(^{12}{\rm C},p)^{23}{\rm Na}$ can lead to a decrease in the amount of C that is available for the $^{20}{\rm Ne}$-creating reaction $^{12}{\rm C}(^{12}{\rm C},\alpha)^{20}{\rm Ne}$, whereas $^{20}{\rm Ne}(\alpha,p)^{23}{\rm Na}$ can directly lead to a decrease in the $^{20}{\rm Ne}$ abundance. Since these reactions are included in the {\tt sagb\_NeNa\_MgAl.net} network but not in {\tt co\_burn\_plus.net}, the former is expected to be more suitable for the computation of realistic SAGB models.
 %We computed models with the same input %physics as in \cite{2018MNRAS.480.1547L}, %including thermohaline  mixing, 
 % adopting both  reaction networks "{\tt %co\_burn\_plus.net}"  \citep[16 species, as %in][] {2018MNRAS.480.1547L} and "{\tt %sagb\_NeNa\_MgAl.net}" (29 species). This %choice is  based on the fact that the %$^{23}$Na isotope is not included in the %"{\tt co\_burn\_plus.net}" network and, in %consequence, neither
 % several  nuclear  reactions  involving  %the  creation  or destruction of $^{20}{\rm %Ne}$ and $^{23}{\rm Na}$ for instance 
%$^{12}{\rm C}(^{12}{\rm C},p)^{23}{\rm Na}$ %and  
%$^{20}{\rm Ne}(\alpha,p)^{23}{\rm Na}$
% among others, but present in the "{\tt %sagb\_NeNa\_MgAl.net}". The latter network %is thus clearly more suitable than the %former for the computation of SAGB models
%  are absent in the "{\tt %co\_burn\_plus.net}" network, but present in %the "{\tt sagb\_NeNa\_MgAl.net}" 
%one  which is thus more suitable for the %computation of SAGB models. 
Extreme mass loss rates were included during the early AGB as in \cite{ 2018MNRAS.480.1547L},  to allow for a detailed comparison with that work. 

The sequence computed with $f=0.1$ and {\tt co\_burn\_plus.net}  thus corresponds to the very same choice of CBM, nuclear network and mass-loss rates adopted by \citet{2018MNRAS.480.1547L}. Consequently, this simulation shows the same abundances reported by \citet{2018MNRAS.480.1547L}, with the formation of $^{20}{\rm Ne}$-dominated WDs (see Fig.~\ref{fig:rednuc}). 
Conversely, and as can be also appreciated in Fig.~\ref{fig:rednuc}, the choice of the larger nuclear network {\tt sagb\_NeNa\_MgAl.net}  leads to a very different  $^{20}{\rm Ne}$ abundance, in better agreement with to those presented by previous authors \citep{1996ApJ...460..489R,2006A&A...448..717S,2010MNRAS.401.1453D,2013ApJ...772...37D}
who find $^{16}$O-dominated instead of $^{20}{\rm Ne}$-dominated cores. It is safe to assume that some of the missing reactions  in {\tt co\_burn\_plus.net} are the key to reproduce the result found by  \citet{2018MNRAS.480.1547L}.
Due to the complex interaction between the production and destruction of isotopes from the  reactions involved, it is a priori not clear which of these reactions are the ones responsible for the enhancement of $^{20}{\rm Ne}$. We accordingly performed an exploration consisting in the one-by-one removal of several specific nuclear reactions and their possible combinations, from the {\tt sagb\_NeNa\_MgAl.net} network that are not present in {\tt co\_burn\_plus.net}, 
which allowed us to look deeper into their impact in the final chemical abundances.

Our results show that it is the removal of the  $^{12}{\rm C}(^{12}{\rm C},p)^{23}$Na reaction what induces the most significant impact on the $^{16}$O and $^{20}$Ne abundances. 
This is shown in Fig.~\ref{fig:no23na}, where we compare the chemical profiles obtained for two sequences, both of which were computed assuming $f=0.1$. In the first one, we assumed the {\tt co\_burn\_plus.net} network. In the second, a modified version of the network {\tt sagb\_NeNa\_MgAl.net}, in which the reaction $^{12}{\rm C}(^{12}{\rm C},p)^{23}{\rm Na}$ was removed, was considered. From these, it is clear that it is the absence of $^{12}{\rm C}(^{12}{\rm C},p)^{23}{\rm Na}$ in  {\tt co\_burn\_plus.net} network what leads to a high $^{20}{\rm Ne}$ composition.

%\begin{table}[]\label{tab:reacs}
%\caption{Nuclear reactions of relevance in the two reaction networks investigated in this paper}
%\begin{tabular}{lll}
%\hline
%\hline
%%reaction $\backslash$ network & {\tt co\_burn\_plus}  & {\tt sagb\_NeNa\_MgAl}    \\
%     &    \multicolumn{2}{c}{Network} \\
%     \cline{2-3} 
%Reaction & {\tt co\_burn\_plus}  & {\tt sagb\_NeNa\_MgAl}    \\ 
%\hline
%$^{16}{\rm O}(\alpha,p)^{19}{\rm F}$&  &   $\checkmark$ \\
%$^{16}{\rm O}(\alpha,\gamma)^{20}{\rm Ne}$ &  $\checkmark$  &    $\checkmark$ \\
%$^{12}{\rm C}(^{12}{\rm C},\alpha)^{20}{\rm Ne}$ &  $\checkmark$ &$\checkmark$  \\
%$^{12}{\rm C}(^{12}{\rm C},p)^{23}{\rm Na}$ &  &$\checkmark$  \\
%$^{20}{\rm Ne}(\alpha,p)^{23}{\rm Na}$ &   $\checkmark$&   $\checkmark$  \\
%$^{23}{\rm Na}(p,\gamma)^{24}{\rm Mg}$ &  $\checkmark$ &    $\checkmark$ \\
%$^{20}{\rm Ne}(\alpha,\gamma)^{24}{\rm Mg}$ &  $\checkmark$ &    $\checkmark$ \\
%$^{24}{\rm Mg}(\alpha,^{12}{\rm C})^{16}{\rm O}$ &  &   $\checkmark$  \\
%$^{20}{\rm Ne}(\alpha,^{12}{\rm C})^{12}{\rm C}$ &  &  $\checkmark$   \\
%$^{20}{\rm Ne}(\gamma,\alpha)^{16}{\rm O}$ &  &  $\checkmark$  \\
%$^{24}{\rm Mg}(\gamma,\alpha)^{20}{\rm Ne}$ &  &  $\checkmark$    \\
%$^{24}{\rm Mg}(\alpha,p)^{27}{\rm Al}$&  &   $\checkmark$  \\  
%\hline
%\end{tabular}
%\end{table}

This result can be understood in terms of the $^{12}$C available during the C-burning phase.
The $^{12}{\rm C}(^{12}{\rm C},p)^{23}$Na nuclear reaction produces the isotope $^{23}$Na  and contributes to the consumption of $^{12}$C. By removing the production of $^{23}$Na,  $^{12}$C is available in the interior of the star and, in overall, allows a larger production of $^{20}$Ne through the reaction $^{12}{\rm C}(^{12}{\rm C},\alpha)^{20}{\rm Ne}$. Consequently about 0.07 of the mass fraction that would have ended as  $^{23}{\rm Na}$  ends up as an additional 0.07 of  $^{20}$Ne by mass fraction. Moreover, for each additional $^{12}{\rm C}(^{12}{\rm C},\alpha)^{20}{\rm Ne}$ we have one free $\alpha$ particle available for reactions, through $^{16}{\rm O}(\alpha,\gamma)^{20}{\rm Ne}$ leading to an additional increase of $^{20}{\rm Ne}$ of about 0.1 (by mass fraction) at the expense of the $^{16}{\rm O}$ abundance. In brief, the removal of the  $^{12}{\rm C}(^{12}{\rm C},p)^{23}{\rm Na}$ nuclear reaction leads to a decrease of about 0.1 in the mass fraction of $^{16}{\rm O}$, the removal of almost all  $^{23}$Na, and an increase of the mass fraction of $^{20}$Ne by about 0.17. 
This is seen in Fig.~\ref{fig:no23na}, where we compare the chemical structure for both models. Clearly,  when the production of $^{23}$Na is removed, the distribution of the most abundant elements $^{20}$Ne and $^{16}$O is in good agreement with those found in \cite{2018MNRAS.480.1547L}.

\section{Consequences for WD evolution}
\label{sect:consequence}

\begin{table}[]
\caption{Effective temperature and \% of crystallized mass for selected values for the three evolutionary sequences computed.}
\resizebox{\columnwidth}{!}{%
\begin{tabular}{llll}
\hline
\hline  
Crystallized mass (\%) & ~~~~~~~~~~~~~$T_{\rm eff}$ (K) \\
\hline
& He-enriched & ONe-core  & NeO-core  \\\cline{2-4}
0 & 30047 &   28714& 30586\\
10&  28272  &27014& 28504\\
20 & 27393&  26027& 27566 \\
30 &  26479  &25282& 26838\\
40 & 25486 &   24360& 25840 \\
50 & 24426  &23253&24747\\
60 & 23069 &   22031&23579\\
70 &  21607  &20665&21989\\
80 & 19748 &    18946 &20043\\
90 & 17012  &16495&17230\\
95 &   14701   &   14229  & 15100\\
96 & 14039& 13696& 14224\\
97 & 13400& 12926& 13401\\
98 & 12358& 12032& 12489\\
99 &  10974&     10758& 11065\\
99.1&10844 & 10593&10945 \\
99.2& 10568& 10336& 10697\\
99.3& 10370& 10163& 10482\\
99.4& 10109& 10000& 10215\\
 \hline
\end{tabular}
}
\label{tab:crist}
\end{table}

In this section we explore the impact of the chemical features  derived from the extreme assumptions discussed in the previous sections on the WD cooling times, crystallization and pulsational properties. 
To this end we computed the evolution of three  $M_{\rm WD}=1.22M_{\odot}$ WD models. We adopted a ONe-core model with the chemical composition derived by \cite{2007A&A...476..893S} consisting of a core  composition of $^{16}$O (56\%),  $^{20}$Ne (29\%) plus traces of $^{12}$C, $^{23}$Na and $^{24}$Mg and a total $^1$H and $^{4}$He content of $1.5\times 10^{-7}$ and $6.5\times 10^{-5}M_{\rm WD}$, respectively. For a better comparison, the H content was set to match the one derived by \citet{2018MNRAS.480.1547L}, while  the He content is the one derived from the correct calculation of the progenitor's evolution.
Additionally, we adopted a NeO-core model obtained from our previous {\tt co\_burn\_plus.net} model, with similar composition derived by \cite{2018MNRAS.480.1547L} that is a core made of $^{16}$O (42\% ), $^{20}$Ne (49\%),  with and H- and He-content added on the top  of 1.5$\times 10^{-7}$ and 7.6$\times 10^{-4}M_{\rm WD}$.
Finally,  to test the impact of the large He content added, we computed a He-enriched model consisting of the same chemical structure as the  correct ONe-core model, but with an incorrect He content of 7.6$\times 10^{-4}M_{\rm WD}$.
 The evolution and structure of the WD models presented in this section were calculated with the  {\tt LPCODE} evolutionary code \citep[for details, see][]{2003A&A...404..593A, 2005A&A...435..631A, 2015A&A...576A...9A, 2021A&A...646A..30A,2016A&A...588A..25M}. During crystallization, we have taken into account the release of latent heat and changes in the core chemical composition resulting from phase separation upon crystallization, using a phase diagram suitable for $^{16}$O and $^{20}$Ne plasmas \citep{2019A&A...625A..87C}.
 
%central compositions
%Reference 56\% 16O, 29\% 20Ne
%test  42\% 16O 49\% 20Ne

\subsection{Cooling times and crystallization}
We computed the cooling times for each WD sequence from $\sim 200\,000$  down to $\sim10\,000$ K. By the time the sequences reach the minimum effective temperature considered, the models are almost completely crystallized.
We found cooling times of $t_{\rm cool}=2.093, 1.882$ and 1.863 Gyr for our ONe-core, He-enriched and NeO-core models respectively.
Our results indicate that the inclusion of large amounts of He induces an acceleration in the cooling process by $\sim 10\%$. 
The main reason behind this result is that, during the enhancement of the He content, heavier elements such as C and O are replaced at the bottom of the He buffer zone by $^{4}$He.
The opacity of $^{4}$He is lower than the one provided by carbon, thus the layers above the degenerate core are less opaque, accelerating the cooling  \citep[see ][]{2021A&A...646A..30A}.
On the other hand, the analysis of the different core compositions reveals that the model with a Ne-dominated core evolves slightly faster. This is because the specific heat per gram of O-rich compositions is larger than in the corresponding Ne-enriched ones. 
%{\bf T cooling (from log teff=5.2 to 4)}

%1)tcooling test LH+PS= 1,863Gyr   (Modelo lauffer)

%2)tcooling reference (Siess + H lauffer) LH+PS=2,093  Gyr (modelo siess 1.22)

%3)tcooling he enriched(+ H lauffer) LH+PS= 1,882Gyr  (modelos siess + he enriched)

Due to the different compositions and cooling rates of our models, some impact is expected on the crystallization degree of the core as the evolution proceeds.
Mixtures enriched in $^{20}$Ne are expected to crystallize earlier than $^{16}$O-rich mixtures. This is seen in Table~\ref{tab:crist}, where we display the percentage of crystallized mass for each of the models computed, along with their corresponding effective temperatures. As can be seen, the NeO-core model starts to crystallize at a temperature that is $\sim 530$~K higher than in the case of our He-enriched model. From the same table,  it is seen that the most important acceleration in the crystallization process is brought about by the large amount of He added,  leading to a crystallization that takes place significantly earlier, i.e., at a temperature that is higher by  of $\sim 1\,300$ K. 
As a consequence, the overall differences in the mass of the solid region can exceed 10\%.
However, as the model cools down, these differences are reduced, and reach only a few percent by the time they reach the  ZZ Ceti instability strip, at $\sim 13\,000$~K. 
%{\color{violet} Under the light of these results we expect that differences in the cooling times and crystallization induced by small changes in the inner core, such as the hybrid structures discussed in  section \ref{sect:CBM}, Fig. \ref{fig:f-todos}  to be of minor relevance in the cooling times and crystallization. }

\subsection{Asteroseismology}

The non-negligible differences found in the size of the crystallized portion of the star and  chemical structure  previously  discussed are expected to leave their imprint on the $g$-mode pulsations of ZZ Ceti stars.  
As is known, the period spectrum and mode-trapping properties of $g$-modes strongly depend on the value of the Brunt-V\"ais\"al\"a frequency across the interior of the star, which in turn, depends on the chemical structure of the model.
  Thus, any change perceived in the chemical profiles are translated into changes in the WD's expected pulsational properties. 
  
  To measure the corresponding impact of the different core compositions and $^4$He content,  we computed the pulsational properties of all our ultra-massive pulsating DA WD models  with the {\tt LP-PUL} pulsation code described in \citet{2006A&A...454..863C}, previously employed in the study of the properties of ultra-massive WD models \citep{2019A&A...621A.100D} and to perform asteroseismologycal studies of ultra-massive ZZ Ceti stars \citep{2019A&A...632A.119C}. Element diffusion was included for all models, from the beginning of the WD cooling track. This process smoothens the inner chemical profiles, strongly affecting the run of the Brunt–V\"ais\"al\"a frequency across the interior.
 We adopted the “hard sphere” boundary conditions when accounting for the effects of crystallization on the pulsational properties of $g$-modes. This assumes that the amplitude of the eigenfunctions of $g$-modes is drastically reduced below the solid/liquid interface, as compared with the amplitude in the fluid region \citep{1999ApJ...526..976M}.

\begin{figure}
  \includegraphics[width=1\columnwidth]{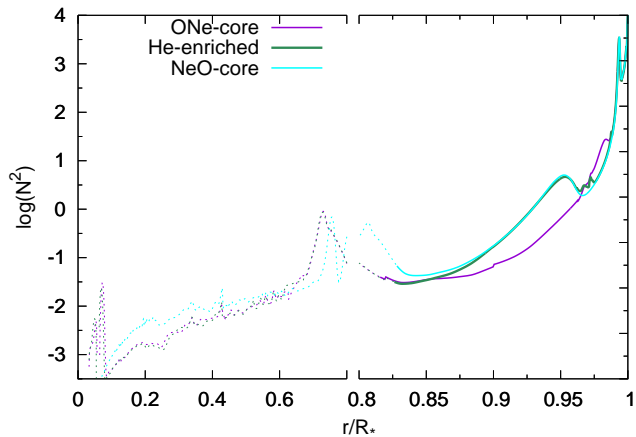}
    \caption{Run of the logarithm of the squared Brunt-V\"ais\"al\"a frequency vs. the normalized radial distance for the three models studied, at $12\, 000$~K. Dotted (solid) lines corresponds to the value of $\log(N^2)$ in the crystallized (fluid) region. Magenta, green, and cyan lines correspond to ONe-rich, He-enriched, and NeO cores, respectively. Note the expanded radial scale on the right side of the plot.}
  \label{fig:bvs}
\end{figure}

 In Fig.~\ref{fig:bvs} we show the run of the logarithm of the squared Brunt-V\"ais\"al\"a frequency ($N^2$) against the normalized radius of the star for each model at $12\,000$ K. The values of the Brunt-V\"ais\"al\"a frequency corresponding to the crystallized region of the model are shown as dotted lines, while the solid lines correspond to the fluid region. Note that the right side of the plot corresponds to a zoomed-in look in the fluid region. At this effective temperature, the models have 98.0\% (ONe-core), 98.3\% (He-enriched), and 98.4\% (NeO-core) of their mass crystallized,  meaning that most of the different core-chemical features do not have an impact on the $g$-mode pulsations. This prevents us from directly measuring the impact of the different core compositions on the pulsational properties.  
 In the fluid region,  however, two features are clearly distinguishable: the bump of the  C/He chemical transition, which is located at $r/R_{\star}\sim 0.95$ for the NeO-core and He-enriched models and at $\sim 0.98$ for the ONe-core model, and the bump of the H/He chemical transition, located at $r/R_{\star}\sim 0.99$.
 The differences found both in the value of $N^2$  at  the C/He chemical transition as well as its location are the result of the artificial addition of $^4$He  above the core. 
 
We computed the asymptotic period spacing, a quantity frequently used in asteroseismological analyses.
As is known,  for  $g$-modes  with  high   radial order $k$,  the separation  of consecutive  periods becomes nearly constant  at a  value  given  by
the  asymptotic  theory of  nonradial stellar  pulsations \citep[e.g.,][and references therein]{2010aste.book.....A, 2015pust.book.....C}.
The asymptotic period spacing is computed as in
\citet{1990ApJS...72..335T}:

\begin{equation} 
\Delta \Pi_{\ell}^{\rm a}= \Pi_0 / \sqrt{\ell(\ell+1)},  
\label{aps}
\end{equation}

\noindent where

\begin{equation}
%\label{asympeq}
\Pi_0= 2 \pi^2 \left[\int_{r_1}^{r_2} \frac{N}{r} dr\right]^{-1},
\label{p0}
\end{equation}

\noindent and $N$ is the  Brunt-V\"ais\"al\"a frequency.

The period spacing thus strongly depends on the value of $N$ and the size of the propagation region of $g$-modes. In fact, when a fraction of the WD core is crystallized, the lower limit of the integral in Eq.~(\ref{p0}) is the radius of the crystallization front, which moves progressively outward as the star cools down and the fraction of crystallized mass increases. As a result, the integral in Eq.~(\ref{p0}) decreases, leading to an increase in the asymptotic period spacing (Eq.~\ref{aps}). 
%Differences in the crystallized mass are translated into differences in the radius of the crystallization front, as can be seen by looking at the starting point of each solid line of Fig.\ref{fig:bvs}, and then affect the value of the integral \ref{p0}.

\begin{figure}
  \includegraphics[width=1\columnwidth]{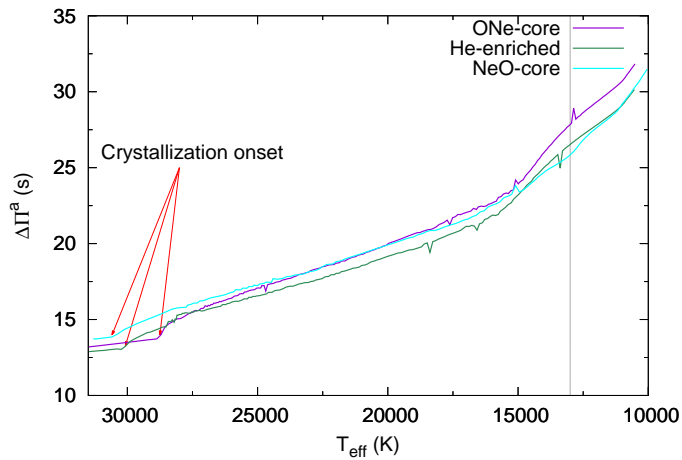}
    \caption{Dipole ($\ell = 1$) asymptotic period spacing as a function of
the effective temperature for the evolutionary cooling sequences of our 
 $M_{\rm WD}= 1.22\, M_{\sun}$ models. The vertical dashed line shows the high-temperature edge  of the ZZ Ceti instability strip. The lines have the same meaning as in Fig.~\ref{fig:bvs}.}
  \label{fig:asymp}
\end{figure}

In Fig.~\ref{fig:asymp} we show the evolution of the asymptotic period spacing as a function of the effective temperature. For each model, a sudden increase in the value of the asymptotic period spacing is noted at the left side of the plot, resulting from the onset of crystallization. From there on, a constant increase of the asymptotic period spacing is seen as the models cool down, mostly due from the outward moving crystallization front.
By the time the models reach the ZZ Ceti instability strip at $\sim 13\,000$ K, the models have very similar crystallized mass ($\sim$ 97\%). Because of this fact, differences found in the asymptotic period spacing are mostly related with differences in the value of the Brunt-V\"ais\"al\"a frequency, i.e., differences induced by the large amount of He  added. This can be seen as an overlap of the 
asymptotic period spacing curves of the NeO-core and He-enriched models.
We found differences in the asymptotic period spacing at this stage of up to 8\%.

Besides crystallization, the most important physical process affecting the shape of the WD structure and, as a consequence, their pulsational properties,  is element diffusion. The chemical structures computed and the results presented in this work reflect thus the effect of element diffusion from the beginning of the WD cooling path until $\sim 10\,000$ K. In particular, we perceived a notable contrast between our derived chemical profile for $^{16}$O with that from \cite{2018MNRAS.480.1547L} who, for models with $1.132\leq M_{\rm WD}/M_{\sun}\leq 1.307$ at $\sim 10\,000$ K, found profiles characterized by a sharp and tall peak at the base of the He-buffer zone.\footnote{ Note that these are exactly the models in which He was added ad hoc to replace the envelope removed by the assumption of artificially strong winds.} As this result can have a significant impact on the trapping properties of pulsating WDs, we did a simple experiment in which we explored how this feature could be produced. 
Being that our SAGB models do not show this signature in the $^{16}$O profile, we assume that it should be formed during the cooling of the WD. 
By assuming that diffusion acts for all the isotopes involved, we were unable to reproduce this peak.  However, we find that such a chemical signature could be produced if the diffusion of the $^{12}$C isotope is ignored.

\section{Conclusions}
\label{sect:concl}
In this work, we have performed several numerical simulations aimed at understanding the discrepant 
core composition of ultra-massive WDs ($M_{\rm WD}>1.05 \, M_\odot$) predicted by previous authors \citep[e.g.][]{1996ApJ...460..489R,2006A&A...448..717S,2013ApJ...772...37D, 2018MNRAS.480.1547L}. Specifically we have addressed the possible formation of Ne-dominated WDs and hybrid CO-Ne WDs and how this can be affected by modeling choices involving for instance the intensity of convective boundary mixing, size of the nuclear reaction network, and extreme winds on the SAGB phase.  We also computed the impact of these 
chemical structures  on the cooling times, crystallization and pulsational properties of ultra-massive ZZ Ceti stars.

Our main conclusions can be enumerated as follows:
\begin{enumerate}
\item The predicted existence of Ne-dominated massive WDs \citep{2018MNRAS.480.1547L} is the consequence of the adoption of a  small nuclear network  ({\tt co\_burn\_plus.net}) which ignores the isotope $^{23}$Na. Specifically, disregarding
the  $^{12}{\rm C}(^{12}{\rm C},p)^{23}{\rm Na}$ nuclear reaction leads to a decrease of about 0.1 in the mass fraction of $^{16}{\rm O}$, the removal of  all  $^{23}{\rm Na}$, and an increase of the mass fraction of $^{20}{\rm Ne}$ by about 0.17 (by mass fraction).
\item In agreement with previous authors, we find that the inclusion of convective boundary mixing quenches the carbon flame, leading to the formation of  hybrid CO-Ne cores. We find that the size of the pure CO core depends both on the adopted value of the overshooting parameter $f$ and the initial mass.
\item Based on the insight coming from 3D hydro-dynamical simulations, we have performed order-of-magnitude estimates which suggest that entrainment at the bottom of the carbon flash convective zone might be of the order of a few tenths of the local pressure scale height. Moreover, given the very slow propagation of the carbon flame in numerical models ($v_{\rm flame} \sim 10^{-3}\,$ cm/s) in comparison with the typical turbulent velocities in the convective zone ($v_{\rm MLT}\sim 5\times 10^4\,$cm/s), it is expected that turbulent entrainment will strongly modify the propagation of the C-flame.

\item The inclusion of strong winds removes the whole H-rich envelope and large part of the He-rich mantle of the star, extinguishing the H-burning shell and affecting the He-burning shell. This has a direct impact in the post-SAGB core's thermal evolution, moderately shifting the location of the temperature maximum. Consequently, the C-flash is ignited closer to the center and the C-burning flame penetrates more deeply. 

\item The prediction by \citet{2018MNRAS.480.1547L} that hybrid CO-Ne cores do not form for massive WDs even when a large amount of convective boundary mixing is included at the bottom of the C-flame is a consequence of the adoption of very large values for the overshooting parameter during the C-flash and the presence of thermohaline mixing.

    \item The inclusion of large amounts of $^{4}$He induces an acceleration in the cooling process of the models by up to $\sim$ 10\%, with respect to those in which the total helium content is obtained by the detailed computation of the SAGB evolution. This result is a  consequence of the lower opacity of helium compared to carbon.
    The adoption of  Ne-rich core mixtures slightly accelerates the cooling process due to the smaller specific heat  per gram, compared to O-rich compositions.
    
    \item Structures composed by  Ne-dominated cores and large amounts of He result in a strong acceleration of the crystallization process, which takes place earlier and thus at an effective temperature up to $\sim 1\,800$~K higher, as compared to our standard models. Differences in the crystallization degree of the models can exceed 10\% for the same effective temperature.
    
    \item At the ZZ Ceti instability strip, differences found in the cooling, crystallization degree and size of the He-buffer zone are expected to impact the asymptotic period spacing 
    by up to $\sim$ 8\%. In addition, the location and size of the base of the He-buffer zone is expected to have a non-negligible impact on the trapping properties of $g$-modes.

\end{enumerate}

These results indicate that, while the formation of hybrid CO-Ne core WDs is a concrete possibility, the existence of Ne-dominated WDs has to be discarded  according to the reaction rates and branching ratios included in the {\tt sagb\_NeNa\_MgAl.net} network.  Additionally, the adoption of these spurious chemical structures for asteroseismological analysis could lead to incorrect determinations of cooling times, crystallization and period spacing's.

%In addition to our main results, we have found during our explorations that {\tt MESA} identifies some of the small convective regions that form by the splitting of the convective core in the late stages of He core burning as "{\tt burn\_z}", and applies convective boundary mixing accordingly. This should be kept in mind when computing evolutionary models that go through a He-core burning phase.

To close, we note that the uncertainties affecting the $^{12}{\rm C}+^{12}{\rm C}$ reaction rate and its  corresponding branching ratios are still very significant \citep[e.g.,][and references therein]{2013ApJ...762...31P,2021ApJ...916...79C, 2021arXiv211115224M}. This could plausibly affect the final chemical profile of the resulting ultra-massive WDs \citep[e.g.,][]{2015AIPC.1645..339H}. A full investigation of this topic is beyond the scope of this paper, but will be addressed by us in the near future

%-----------------------------------------------------------------
\begin{acknowledgements}
FCDG acknowledges financial support provided by FONDECYT grant 3200628. This project was supported in part by ANID's Millennium Science Initiative through grant ICN12\textunderscore 12009, awarded to the Millennium Institute of Astrophysics (MAS), and by Proyecto Basal AFB-170002.
Part of this work was supported by PIP 112-200801-00940 grant
from CONICET, grant G149 from University of La Plata, and
grant PICT 2016-0053 from ANPCyT, Argentina. FP is supported by a EVC-CIN fellowship for undergraduate students. This research has made use of
NASA Astrophysics Data System.
\end{acknowledgements}

%   \bibliography{Yourfile.bib} % your references Yourfile.bib
\bibliography{fran}
\bibliographystyle{aa}

{\it Software:} Modules for Experiments in Stellar Astrophysics
\citep[MESA;][]{Paxton2011, Paxton2013, Paxton2015, Paxton2018, Paxton2019}, {\tt MESASDK} 20.3.1 \citep{2020zndo...3706650T}, {\tt mkipp} \citep{2019zndo...2602098M}. {\tt LPCODE}, \citep{2021A&A...646A..30A}.

%\begin{appendix}\label{app:mesa} %First appendix
%\section{MESA Input Physics}
%The MESA equation of state (EOS) is a blend of the OPAL \citep{Rogers2002}, SCVH
%\citep{Saumon1995}, FreeEOS \citep{Irwin2004}, HELM \citep{Timmes2000},
%PC \citep{Potekhin2010}, and Skye \citep{Jermyn2021} EOSes.

%Radiative opacities are primarily from OPAL \citep{Iglesias1993,
%Iglesias1996}, with low-temperature data from \citet{Ferguson2005}
%and the high-temperature, Compton-scattering dominated regime by
%\citet{Poutanen2017}.  Electron conduction opacities are from
%\citet{Cassisi2007}.

%Nuclear reaction rates are from JINA REACLIB \citep{Cyburt2010} plus
%additional tabulated weak reaction rates \citep{Fuller1985, Oda1994,
%Langanke2000}.  Screening is included via the prescription of \citet{Chugunov2007}.
%Thermal neutrino loss rates are from \citet{Itoh1996}.
%\end{appendix}

%\end{thebibliography}

%\begin{appendix} %First appendix
%\section{Nuclear reaction nets}
%      \subsection{"co\_burn\_plus"}

%\subsection{"sagb\_NeNa\_MgAl}

\end{document}